**Quantifying cancer epithelial-mesenchymal plasticity and its association with stemness and immune response**

Dongya Jia[1], Xuefei Li[1], Federico Bocci[1,2], Shubham Tripathi[3], Youyuan Deng[1,4], Mohit Kumar Jolly[5,*], José N. Onuchic[1,2,6,7,*], Herbert Levine[8,9,*]

[1]Center for Theoretical Biological Physics, Rice University, Houston, TX 77005, USA
[2]Department of Chemistry, Rice University, Houston, TX 77005, USA
[3]PhD Program in Systems, Synthetic, and Physical Biology, Rice University, Houston, TX 77005, USA
[4]PhD Program in Applied Physics, Rice University, Houston, TX 77005, USA
[5]Centre for BioSystems Science and Engineering, Indian Institute of Science, Bangalore 560012, India
[6]Department of Biosciences, Rice University, Houston, TX 77005, USA
[7]Department of Physics and Astronomy, Rice University, Houston, TX 77005, USA
[8]Department of Bioengineering, Northeastern University, Boston, MA 02115, USA
[9]Department of Physics, Northeastern University, Boston, MA 02115, USA

*Correspondence:
Mohit Kumar Jolly: mkjolly@iisc.ac.in
José N. Onuchic: jonuchic@rice.edu
Herbert Levine: h.levine@northeastern.edu

**Abstract**

Cancer cells can acquire a spectrum of stable hybrid epithelial/mesenchymal (E/M) states during epithelial-mesenchymal transition (EMT). Cells in these hybrid E/M phenotypes often combine epithelial and mesenchymal features and tend to migrate collectively commonly as small clusters. Such collectively migrating cancer cells play a pivotal role in seeding metastases and their presence in cancer patients indicates an adverse prognostic factor. Moreover, cancer cells in hybrid E/M phenotypes tend to be more associated with stemness which endows them with tumor-initiation ability and therapy resistance. Most recently, cells undergoing EMT have been shown to promote immune suppression for better survival. A systematic understanding of the emergence of hybrid E/M phenotypes and the connection of EMT with stemness and immune suppression would contribute to more effective therapeutic strategies. In this review, we first discuss recent efforts combining theoretical and experimental approaches to elucidate mechanisms underlying EMT multi-stability (i.e. the existence of multiple stable phenotypes during EMT) and the properties of hybrid E/M phenotypes. Following we discuss non-cell-autonomous regulation of EMT by cell cooperation and extracellular matrix. Afterwards, we discuss various metrics that can be used to quantify EMT spectrum. We further describe possible mechanisms underlying the formation of clusters of circulating tumor cells. Last but not least, we summarize recent systems biology analysis of the role of EMT in the acquisition of stemness and immune suppression.

**Introduction**

Metastasis remains the major cause of cancer-related deaths [1]. To enable successful metastasis, cancer cells often engage a trans-differentiation program referred to as epithelial-mesenchymal transition (EMT) in order to promote migratory and invasive properties [2]. During EMT, cells gradually lose epithelial features such as a cobblestone-like morphology, cell-cell adhesion and apico-basal polarity and acquire mesenchymal features such as a spindle-like morphology, increased motility and invasiveness [2]. The concept of EMT was initially described during embryonic development. EMT was first observed *in vitro* by Greenberg and Hay showing that epithelial cells suspended in three-dimensional collagen gels lose their apical-basal polarity and acquire characteristics of migrating mesenchymal cells [3]. Later *in vivo* work by Nieto *et al.* argued that EMT is essential for the formation of mesoderm and the generation of the migratory neutral crest cells during chicken embryonic development [4]. EMT also plays a critical role during physiological wound repair and pathological fibrosis [5]. In the context of cancer metastasis, EMT has been proposed to be typically associated with enhanced metastatic potential of cancer cells [2] and the reverse process - mesenchymal-epithelial transition (MET) – has been considered to facilitate effective metastatic colonization by regaining epithelial and proliferative traits that are lost during EMT [6].

During metastasis of tumors, cells rarely undergo a complete EMT and enter a fully mesenchymal phenotype [7]. Instead, partial EMT leading to a hybrid epithelial/mesenchymal (E/M) phenotype has often been observed [8]. In other words, recent studies have emphasized that contrary to the prevailing dogma of EMT being a binary process, there exists a spectrum of hybrid E/M phenotypes characterized by varying extents of epithelial and mesenchymal features and associated with metastatic potential and invasiveness [9,10]. Cancer cells in these hybrid E/M phenotypes tend to combine epithelial (e.g. cell-cell adhesion) and mesenchymal (e.g. increase motility) traits [11] and can thus migrate collectively as a cluster, which can reach and enter the bloodstream intact. Clusters of circulating tumor cells (CTCs) or CTC clusters contribute much more than their proportional share to forming metastases relative to individual CTCs which are typically mesenchymal cells [12]. Clinical evidence supports the aggressiveness of CTC clusters and their presence in the peripheral blood has been shown to be a prognostic factor for poor patient survival across multiple types of cancer [13]. Therefore, a better understanding of the mechanisms underlying the emergence of hybrid E/M phenotypes and the formation of CTC clusters can lead to more effective therapeutic designs targeting metastasis.

Recently there have been ongoing debates regarding the necessity of EMT for metastasis [14]. Zheng *et al.* demonstrated that deletion of the EMT-inducing transcription factor (EMT-TF) SNAIL or TWIST in genetically engineered mouse models of pancreatic ductal adenocarcinoma (PDAC) did not cause any significant change in tumor progression and metastasis [15]. Fischer *et al.* suggested that EMT inhibition by over-expression of microRNA (miR)-200 led to no obvious change in lung metastasis development in spontaneous breast-to-lung metastasis mouse models [16]. From our perspective, these two studies assumed that EMT can be completely repressed by single factor manipulation, for example, deletion of SNAIL or TWIST in PDAC or over-expression of miR-200 in breast cancer. Following these studies, Krebs *et al.* used the same PDAC model [15] and showed that deletion of the EMT-TF ZEB1 significantly suppresses the colonization capacity of tumor cells and the formation of metastases [17], indicating a non-redundant role of EMT-TFs in regulating PDAC metastasis [18]. Moreover, another study by Cursons *et al.* showed that overexpression of miR-200c in the HMLE-derived mesenchymal cells established by exposing HMLE cells to TGF-β for 24 days can only drive a partial MET where the canonical epithelial marker E-cadherin increases but the mesenchymal maker vimentin remains [19]. One major contributing factor to these debates regarding the functional role of EMT in metastasis is the lack of consistency in defining EMT itself, owing to its highly nonlinear and multidimensional nature [20]. Thus, a rigorous quantification of the EMT status of tumor cells and a systematic analysis of the interacting EMT regulators such as interactions between miRNAs, EMT-TFs and epigenetic factors is urgently needed.

Notably, various extracellular biochemical and biomechanical factors can govern the induction and maintenance of a partial or complete EMT. For instance, neighboring cells can induce EMT through TGF-β secretion [21] or Notch signaling [22], and also the alteration of stiffness of extracellular matrix (ECM) can trigger EMT [23]. In addition to the complexity of EMT itself, cancer cells undergoing EMT tend to acquire "stemness" characteristics, which are believed to be responsible for tumor-initiation ability and therapy resistance [24]. Moreover, cancer cells can interact with many other types of cells in the tumor microenvironment, such as fibroblasts, endothelial cells and immune cells [25,26]. Specifically, the co-evolution of cancer and immune cells [27,28] has captured attentions recently due to the promising effects of cancer immunotherapy [29,30]. Because of the typically enhanced metastatic potential and therapy resistance of cancer cells undergoing EMT, it is natural to analyze the correlation and causal relationship between EMT and immune response [31,32]. Furthermore, both EMT [1,2] and immune signatures [33–39] have been shown to be prognostic indicators for various types of cancer. A better understanding of their relationship can potentially contribute to more effective therapeutic designs.

In this review, we focus on how a combination of theoretical and experimental efforts has led to a better understanding of the EMT multi-stability, the characteristics of the hybrid E/M phenotypes and the association of EMT with the acquisition of stemness and immune suppression (**Figure 1**). We start with a discussion of mathematical modeling studies of EMT regulatory networks, that elucidate mechanisms underlying EMT multi-stability and particularly the emergence of hybrid E/M phenotypes. Then we summarize recent *in vitro* and in *vivo* experimental studies that characterize hybrid E/M phenotypes. We then discuss non-cell-autonomous regulation of EMT by cell cooperation and ECM. We further discuss metrics that have been developed to quantify EMT status without pre-supposing that cells lie at the extreme ends of completely epithelial or completely mesenchymal. Finally, we extend our discussion to the coupling of EMT with stemness and immune response.

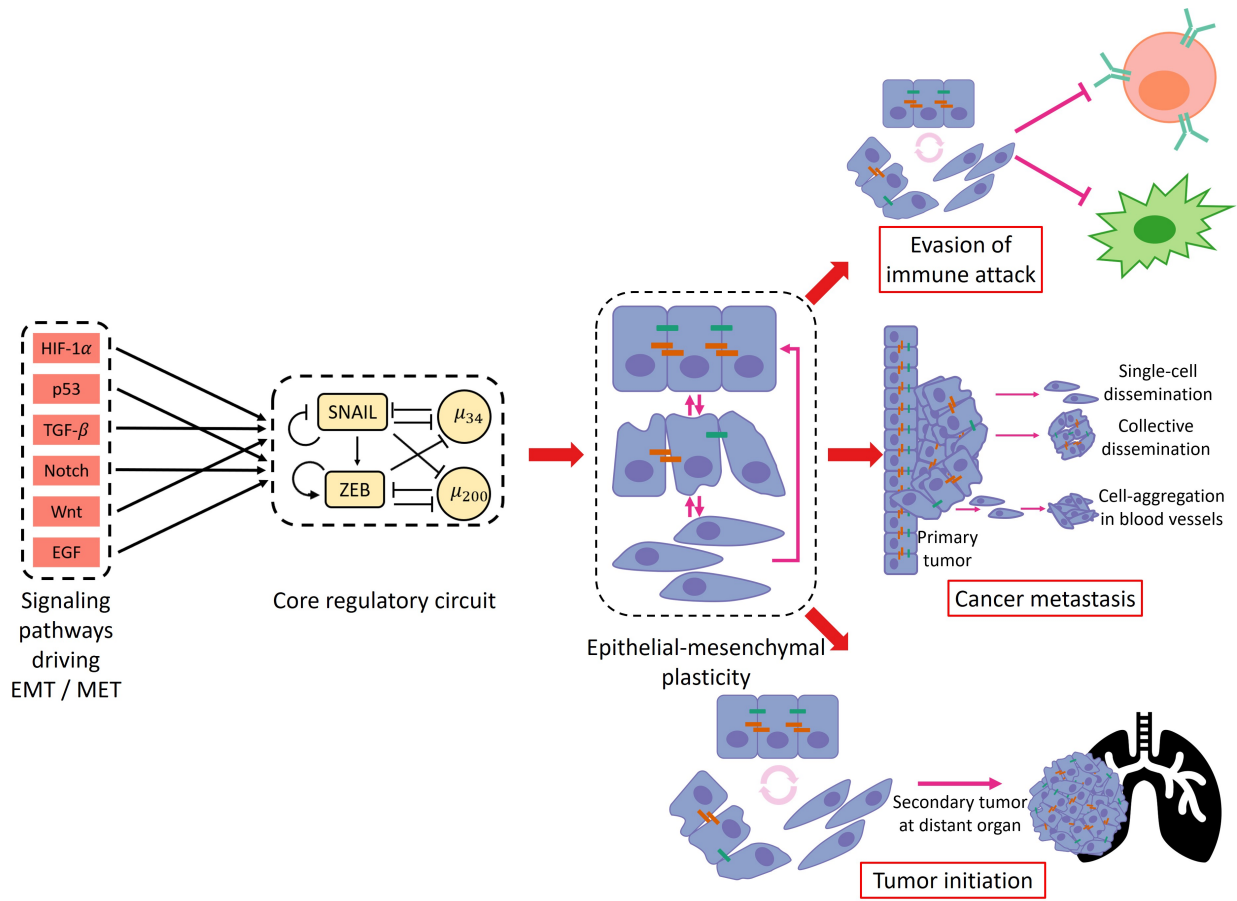

**Figure 1. An overview of epithelial-mesenchymal plasticity - causes and consequences**. Multiple signaling pathways driving EMT/MET tend to converge to a core regulatory circuit which involves both transcriptional and post-transcriptional regulation. The core regulatory circuit exhibits multi-stable dynamics and govern the transitions between different EMT phenotypes. The epithelial-mesenchymal plasticity enables both collective and individual cell invasion and has been implicated in multiple processes including avoiding immune destruction and heightened tumor initiation.

**Emergence of hybrid epithelial/mesenchymal phenotypes**

*Hybrid E/M phenotypes are predicted by mathematical modeling of EMT regulation*

EMT is governed by a complex gene regulatory network (GRN) including miRNAs, transcription factors (TFs), alternative spicing factors, epigenetic modifiers, growth factors, long non-coding RNAs and others [7,40,41]. Several groups have proposed that two microRNA families miR-200 and miR-34 interacting with two EMT-TF families ZEB and SNAIL tend to form a core EMT regulatory network [40]. Many signaling pathways such as TGF-β, WNT and Notch impinge upon this network to regulate EMT. The miR-200 and miR-34 function as guardians of the epithelial phenotype and ZEB and SNAIL promote EMT. Mechanism-based mathematical modeling of this network that includes a detailed treatment of microRNA-mediated regulation suggests that it can give rise to three stable states: an epithelial phenotype characterized by miR-200$^{high}$/ZEB$^{low}$/miR-34$^{high}$/SNAIL$^{low}$; a mesenchymal phenotype characterized by miR-200$^{low}$/ZEB$^{high}$/miR-34$^{low}$/SNAIL$^{high}$; and a hybrid E/M phenotype characterized by co-expression of miR-200 and ZEB [42]. According to this model, the miR-200/ZEB circuit can function as a three-way decision-making switch governing the transitions between epithelial, mesenchymal and hybrid E/M phenotypes and the miR-34/SNAIL circuit primarily functions as a noise-buffering integrator [42]. Alternatively, a different characterization of the hybrid E/M state has been proposed: starting from an epithelial state - miR-200$^{high}$/ZEB$^{low}$/miR-34$^{high}$/SNAIL$^{low}$, a hybrid state can be achieved when the miR-34/SNAIL circuit switches from miR-34$^{high}$/SNAIL$^{low}$ to miR-34$^{low}$/SNAIL$^{high}$, but the miR-200/ZEB circuit is maintained at miR-200$^{high}$/ZEB$^{low}$ [43]. Despite these differences [44], both of these mathematical models clearly indicate that EMT need not be a binary process and instead a stable hybrid E/M state expressing both epithelial and mesenchymal traits can be the end point of a transition.

The existence of hybrid E/M states has been further supported by other computational studies analyzing extended versions of the core EMT regulatory network [45–47]. Steinway *et al.* showed combinatorial intervention of TGF-β signal and SMAD suppression can lead to multiple hybrid E/M states using Boolean modeling [45]. Huang *et al.* and Font-Clos *et al.* showed that the hybrid E/M phenotypes are robust stable states emerging due to the topologies of EMT regulatory networks [46,48–50]. Mathematical modeling approaches have been further used to characterize phenotypic stability factors (PSFs) that can promote and stabilize hybrid E/M states. These PSFs include the transcription factors OVOL, GRHL2, NRF2, ΔNP63α, NUMB and miR-145/OCT4 [50–54]. These PSFs can function in two related manners. First, coupling these PSFs with the decision-making circuit of EMT - miR-200/ZEB expands the parameter space and thereby the expected physiological conditions under which a hybrid E/M state can be attained [51–53]. In particular, PSF coupling can create a region of parameter space in which the only stable state is a hybrid one. Second, these PSFs increase the mean residence time of the hybrid E/M state, and thus its expected percentage in a cell population [50]. Experimental validation for these PSFs comes from observations that knocking them down destabilizes hybrid E/M phenotype and collective cell migration, instead promoting a complete EMT and individual cell migration. Other mechanisms through which a stable hybrid E/M phenotype can be acquired rely on combinatorial treatments with EMT-inducing and MET-inducing signals [54,55] or an increase of gene expression noise [56].

In summary, these mathematical modeling studies provide insights into the multi-stable nature of EMT, particularly the existence and characterization of hybrid E/M states. As we will now see, these modeling-predicted hybrid E/M states have been recently characterized experimentally both *in vitro* and *in vivo*.

*In vitro characterization of hybrid E/M phenotypes*

To map the EMT spectrum in ovarian carcinoma, Huang *et al.* analyzed the protein levels of epithelial markers – E-cadherin (E-Cad) and pan-cytokeratin (Pan-CK) and the mesenchymal marker – vimentin (Vim) of 42 ovarian carcinoma cell lines. Among these 42 cell lines, 9 epithelial cell lines are characterized by E-Cad$^+$/Pan-CK$^+$/Vim$^-$, 7 mesenchymal cell lines are characterized by E-Cad$^-$/Pan-CK$^-$/Vim$^+$, and 26 hybrid E/M cell lines are characterized by either E-Cad$^+$/Pan-CK$^+$/Vim$^+$ (n=18, referred to as intermediate E) or E-Cad$^-$/Pan-CK$^+$/Vim$^-$ (n=8, referred to as intermediate M) [57]. The intermediate E ovarian carcinoma cell lines exhibit significantly higher levels of SNAI1 mRNA and lower levels of ZEB1/2 mRNAs relative to the intermediate M ovarian carcinoma cell lines. The different expression patterns of SNAI1 and ZEB1/2 in intermediate E and intermediate M is reminiscent of the different characterizations of the hybrid E/M states by Lu *et al.* [42] and Zhang *et al.* [43]. Interestingly, the intermediate M ovarian carcinoma cell line SKOV3 exhibited significantly higher spheroidogenic efficiency, migratory and invasive potential relative to the ovarian carcinoma cell lines with other phenotypes. Further studies have revealed underlying feedback loops that can regulate such phenotypic plasticity in ovarian cancer [58].

Similarly, to characterize the EMT spectrum in lung adenocarcinoma, Schliekelman *et al.* analyzed the cell morphologies and the ratios of surface localized E-cadherin to vimentin of 38 non-small cell lung cancer (NSCLC) cell lines out of which 9 were binned as epithelial, 9 as mesenchymal and 20 as hybrid E/M [59]. Notably, in these experiments the hybrid E/M cell lines are identified at a population level and therefore can contain purely individually hybrid E/M cells, or alternatively contain a mixture of epithelial and mesenchymal cells or both. Among these hybrid

E/M NSCLC cell lines, almost all individual H1975 cells were shown to stably co-express E-cadherin and vimentin at least for two months over multiple passages, thus representing stable hybrid E/M cells [51]. In contrast, individual NSCLC H1944 or H2291 cells express either only E-cadherin or only vimentin, thus these cell lines are largely a mixture of epithelial and mesenchymal cells [60,61]. When knocking down the predicted PSFs - GRHL2, OVOL2, NUMB or NRF2 - via siRNAs in H1975 cells, these hybrid E/M cells transition to a complete mesenchymal phenotype [51,53,62]. Another cell line that exhibits hybrid E/M phenotype characterized by co-expression of E-cadherin and ZEB1 at a single-cell level is human bladder cancer (HBC) RT4. Overexpression of the PSF NRF2 in RT4 cells increases the protein levels of both E-cadherin and ZEB1, supporting the predicted role of NRF2 in stabilizing a hybrid E/M phenotype.

In addition to cell lines containing either individual hybrid E/M cells or a mixture of E and M cells, there are cell lines that exhibit co-existence of hybrid E/M cells together with epithelial and/or mesenchymal cells. For example, the NSCLC HCC827 cells contain mostly epithelial cells and a subpopulation of individual hybrid E/M cells characterized by co-expression of epithelial markers including E-cadherin and miR-200a/b/c and mesenchymal markers including vimentin, ZEB1 and SNAI1 [63]. Treatment of the HCC827 cells with the epidermal growth factor receptor (EGFR) inhibitor erlotinib induces a stably erlotinib-resistant cell population among which the percentage of hybrid E/M cells is increased relative to their parental erlotinib-sensitive HCC827 cells [63], indicating a correlation of hybrid E/M phenotypes with therapy resistance. As expected, these HCC827-derived erlotinib-resistant cells exhibit collective migration and form more spheroids relative to their parental erlotinib-sensitive HCC827 cells [63]. Other NSCLC cell lines such as H920 and H2228 have been shown to be mixtures of hybrid E/M and epithelial cells with the hybrid E/M ones being dominant [60]. Aside from NSCLC cells, Grosse-Wilde *et al.* used flow cytometry analysis to isolate a subpopulation of breast cancer HMLER cells that are characterized by $CD24^+/CD44^+$. Most of these $CD24^+/CD44^+$ HMLER cells co-express epithelial genes such as CDH1 and EPCAM and mesenchymal genes such as VIM and ZEB2 and thus exist in a hybrid E/M phenotype [64]. These hybrid E/M HMLER cells demonstrate maximum mammosphere-forming ability relative to their epithelial and mesenchymal counterparts, highlighting that a hybrid E/M phenotype will often be more aggressive than a complete EMT phenotype, a theme which has also found emerging clinical support [65]. Later on, we will discuss how these additional properties of hybrid cells may arise due a coupling of the EMT pathway and the network determining "stemness".

As already noted, hybrid E/M phenotypes can be acquired and maintained by combinatorial treatments of EMT- and MET-inducing signals. For example, Gould *et al.* showed that the epithelial colon carcinoma DLD1 cells can undergoing a partial EMT and acquire a hybrid E/M phenotype co-expressing E-cadherin and vimentin. Such hybrid E/M DLD1 cells are driven by simultaneous expression of the TFs - pSP1 and NFATc in response to the combined treatment of VEGF-A and TGF-β1/2 [54]. Similarly, Biddle *et al.* showed that treatment of the oral squamous cell carcinoma (OSCC) CA1 and LM cells with TGF-β and retinoic acid simultaneously can stabilize a hybrid E/M subpopulation characterized by $CD44^{high}/EpCAM^{low/-}/CD24^{+}$ [55]. Additional experimental studies characterizing hybrid E/M phenotypes and their implications have been reviewed elsewhere [66,67]. In summary, hybrid E/M phenotypes have been observed in *vitro* at a single-cell level across multiple cancer types.

The co-existence of epithelial, mesenchymal and hybrid E/M subpopulations in a single cell line indicates a population heterogeneity of EMT. Such heterogeneity can be generated and maintained via multiple mechanisms all of which can contribute to the acquisition and maintenance of hybrid E/M phenotypes (**Figure 2**). First, the EMT regulatory networks can be multi-stable [42,43,46,48]. Noise in the expression levels of the involved RNAs and proteins can thus cause transitions from one stable steady state to another [68]. Such noise may arise from the inherent stochasticity of the transcription process in cells [69] or from the random partitioning of parent cell RNAs and proteins among the daughter cells at the time of cell division [70–72]. Since both noise sources are cell-autonomous,, individual epithelial cells may spontaneously undergo a phenotypic transition and acquire (or give rise to, in the second scenario) a hybrid E/M phenotype [73]. Thus, one must be careful when choosing a cell line for experiments - a cell line known to be epithelial may include substantial fractions of hybrid E/M and/or mesenchymal cells [55,73]. Second, the population heterogeneity of EMT can arise via cell-cell communication. An example of such a communication channel is Notch-Delta-Jagged signaling. Some cells with high levels of Delta/Jagged expression act as senders, while other cells with high levels of Notch receptor expression act as receivers.

This leads to a population that is inherently heterogeneous - a mix of sender and receiver cells. Notch-Delta-Jagged signaling-mediated heterogeneity is closely tied to the emergence of hybrid E/M phenotypes [74]. This will be discussed in more detail below. Third, population heterogeneity of EMT can arise due to different kinetic parameters controlling gene regulation in different cells. Due to the multitude of peripheral factors involved in governing the behavior of a core regulatory circuit, the kinetics of the core circuit can vary from cell to cell leading to different responses to the same external cues in different cells [46]. This might be connected to the aforementioned random partitioning but might arise as well due to different long-lived fluctuations, perhaps related to chromatic structure heterogeneity.

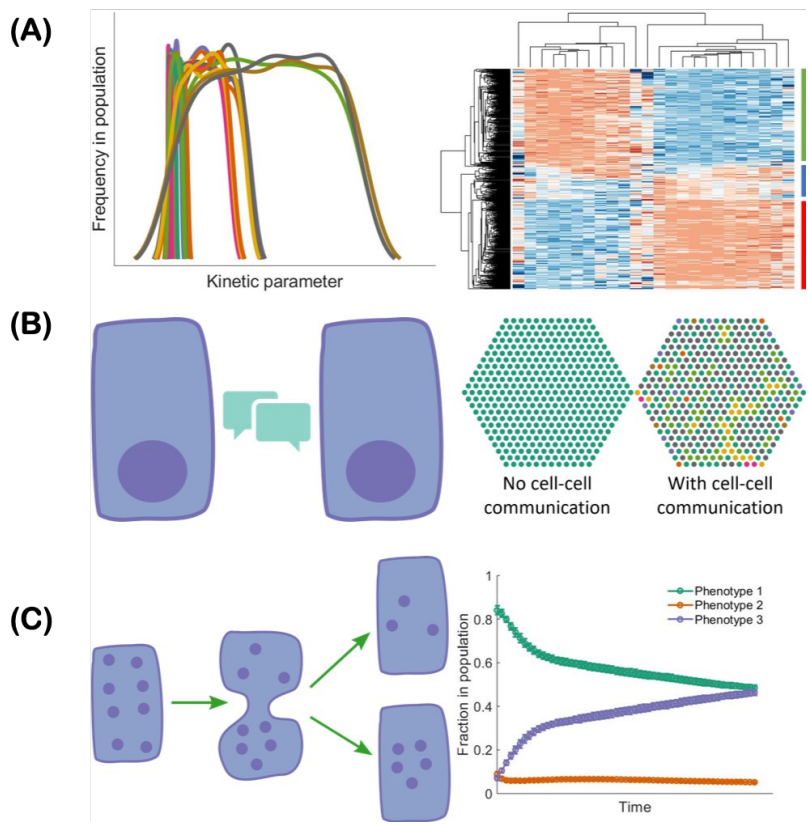

**Figure 2. Multiple mechanisms can lead to a population heterogeneity of EMT.** (A) The kinetics of gene regulation can vary in different cells due to stochasticity in kinetic parameters. Thus, each parameter can exhibit some variation in its value in a given isogenic population (left: each curve denotes the distribution of values of an individual parameter in a given population). This heterogeneity can lead to cells in the population acquiring different phenotypes in response to the same external cues (right: heatmap indicates the emergence of various phenotypes. Each row represents a cell and each column represents expression of one gene. Red color represents relatively high gene expression and blue color represents relatively low gene expression). (B) Cell-

cell communication (left) mediated by, for example Notch-Delta-Jagged signaling, can lead to cells spontaneously acquiring "signal sender", "signal receiver" and "signal sender/receiver" phenotypes in a population (right, different colors correspond to different cell phenotypes). (C) During cell division, noise can lead to asymmetric partitioning of molecules among daughter cells, thus leading to various phenotypes of daughter cells. As a result, the fractions of cells exhibiting different phenotypes can change over time.

*In vivo characterization of hybrid E/M phenotypes*

To identify whether cancer cells can acquire hybrid E/M phenotypes *in vivo*, Pastushenko *et al.* used fluorescence-activated cell sorting (FACS) to screen cell surface markers of skin squamous cell carcinoma (SCC) cells that can undergo spontaneous EMT and generate EpCAM$^+$ epithelial cells and EpCAM$^-$ mesenchymal-like cells in genetically engineered mouse models [10]. While the EpCAM$^+$ cells exhibit homogeneous expression of most of the markers, the EpCAM$^-$ cells exhibit heterogeneous expression of 17 cell surface markers among which the most heterogeneously expressed are CD61, CD51 and CD106. Consequently, the authors used combinatorial multicolor FACS analysis of these three markers to further classify the EpCAM$^-$ cells into 6 subpopulations among which the CD51$^-$CD61$^-$CD106$^-$, CD51$^+$ and CD106$^+$ subpopulations exhibit co-expression of the epithelial marker keratin 14 (K14) and the mesenchymal marker vimentin, thus can be tentatively considered to be hybrid E/M phenotypes. The CD51$^+$CD61$^+$CD106$^+$ and CD51$^+$CD61$^+$ subpopulations, on the other hand, are vimentin positive and K14 negative and thus being mesenchymal-like. Intriguingly, the hybrid E/M CD51$^-$CD61$^-$CD106$^-$ and CD106$^+$ subpopulations generate significantly more metastases relative to other subpopulations, though all subpopulations here share similar tumour-propagating capacity. Similar results have also been observed in MMTV-PyMT mammary luminal tumours [10]. Another study performed by Aiello *et al.* used a lineage-tracing mouse model of PDAC showed that the PDAC cells undergoing EMT can acquire a hybrid E/M phenotype via re-localization of epithelial proteins such as E-cadherin and claudin-7 from membrane to intracellular foci [75]. The emergence of these hybrid E/M PDAC cells indicates that besides transcriptional control, post-transcriptional regulation of localization can be also important to mediate the existence of hybrid E/M phenotypes. Finally, another study by Puram *et al.* showed that the head and neck squamous cell carcinoma (HNSCC) cells from patients exhibit a hybrid E/M phenotype characterized by co-expression of epithelial markers such as EPCAM and KRT17 and mesenchymal markers such as vimentin and TGF-β-induced (TGFBI) through single-cell transcriptomic analysis [76]. Intriguingly, these hybrid E/M HNSCC cells tend to localize at the leading edge of tumors close to surrounding stroma cells.

In summary, mathematical modeling together with both *in vitro* and *in vivo* experimental studies consistently demonstrate the existence of multiple hybrid E/M phenotypes characterized by varying extents of epithelial and mesenchymal features (**Figure 3**). It is worth noting that in addition to the characterization of the hybrid E/M phenotypes, mathematical modeling of EMT regulatory networks have generated many other interesting predictions that have been recently validated by experimental studies. First, modeling the core EMT regulatory network – miR-200/ZEB/miR-34/SNAIL suggests a sequential response of the EMT-TFs SNAIL and ZEB to the EMT inducing signal TGF-β [43,44]. When treated with different levels of TGF-β, SNAIL is predicted to precede ZEB and to be upregulated at relatively low TGF-β levels. The predicted different responses of SNAIL and ZEB has been verified in MCF10A cells [43]. Second, another prediction from modeling studies [42,43] is that EMT and MET are not necessarily symmetric processes and hysteresis is expected during EMT. The predicted hysteretic behavior of EMT has been recently demonstrated during TGF-β induced EMT of NMuMG and EpRAS cells, where the levels of E-cadherin exhibit a bimodal transition. Such hysteretic behavior of E-cadherin is regulated by the miR-200/ZEB1 circuit and blocking the inhibition of miR-200 by ZEB1 (referred to in this study as mutant cells) results in only a unimodal transition of E-cadherin, though these mutant cells can still undergo EMT with changes of EMT markers at a similar degree relative to the wild type [77]. Moreover, TGF-β induced EMT of mutant cells exhibits significantly decreased sphere-formation ability *in vitro*, decreased frequency of tumor-initiating cells and lung metastases *in vivo* relative to their wild types. These results also confirm a prominent role of the miR-200/ZEB circuit in regulating the aspects of EMT dynamics that result in a variety of functional consequences.

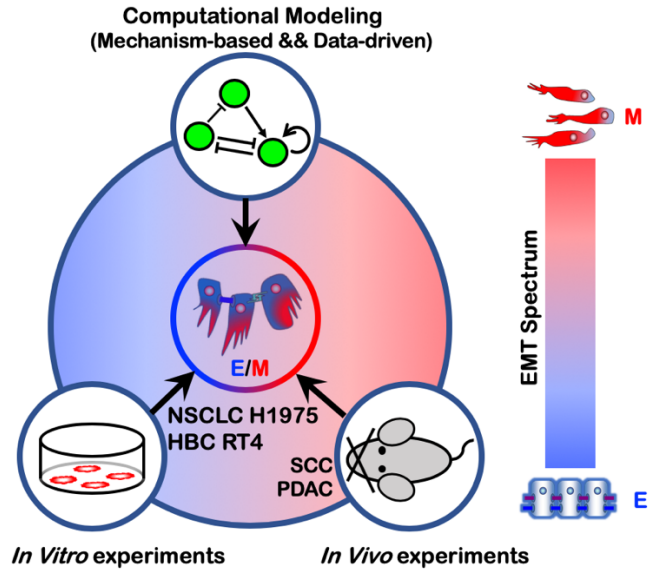

**Figure 3. Emergence of hybrid E/M phenotypes demonstrated by a combination of theoretical and experimental efforts.** The cell lines NSCLC H1975 and HBC RT4 exhibit a hybrid E/M phenotype at a single-cell level. Hybrid E/M phenotypes have also been characterized *in vivo* using mouse models of SCC and PDAC.

**EMT regulation by cell cooperation**

*Notch signaling*

Although EMT is fundamentally an individual cell phenomenon, it can also be regulated by cell cooperation such as Notch signaling. Notch signaling is a cell-cell communication mechanism and highly conserved across species. Originally characterized in Drosophila development, Notch signaling is a conserved and well-known regulator of multiple hallmarks of cancer, including angiogenesis and EMT [78–80].

The Notch signaling cascade is initiated by the binding of the Notch transmembrane receptor with a ligand belonging to a neighboring cell. This binding leads to the cleavage of the Notch intracellular domain (NICD), which is then released in the cytoplasm and transported to the cell nucleus, where it acts as a transcriptional cofactor [81]. Notch signaling is deeply coupled to the EMT regulatory networks discussed in previous sections. For example, on one hand, EMT-inhibiting miRNAs miR-34 and miR-200 reduce the levels of Notch receptor and ligands [82–84] by translational regulation [85]. On the other hand, NICD promotes the transcription of SNAIL and thus acts as an EMT inducer [86,87]. Therefore, cancer cells undergoing EMT can in turn induce EMT in their neighboring cells through the binding of their ligands to a neighbor's Notch receptors.

As often seen in the developmental context, Notch signaling can give rise to different spatial patterns of cell phenotypes due to feedback regulation between NICD and various alternate ligands. Specifically, NICD transcriptionally represses ligands of the Delta family but activates ligands of the Jagged family. Therefore, signaling through the Notch-Delta pathway typically promotes opposite cell fates in neighboring cells, or 'lateral inhibition'. This is accomplished by amplifying initial differences in the levels of receptors and ligands, ultimately leading to one cell with high levels of Notch receptor and low levels of Delta ligand (receiver cell) and a neighbor cell with low Notch and high Delta (sender cell) [88,89]. Conversely, Notch-Jagged signaling typically promotes a similar cell fate in neighboring cells, or 'lateral induction', because NICD upregulates Jagged ligands [90].

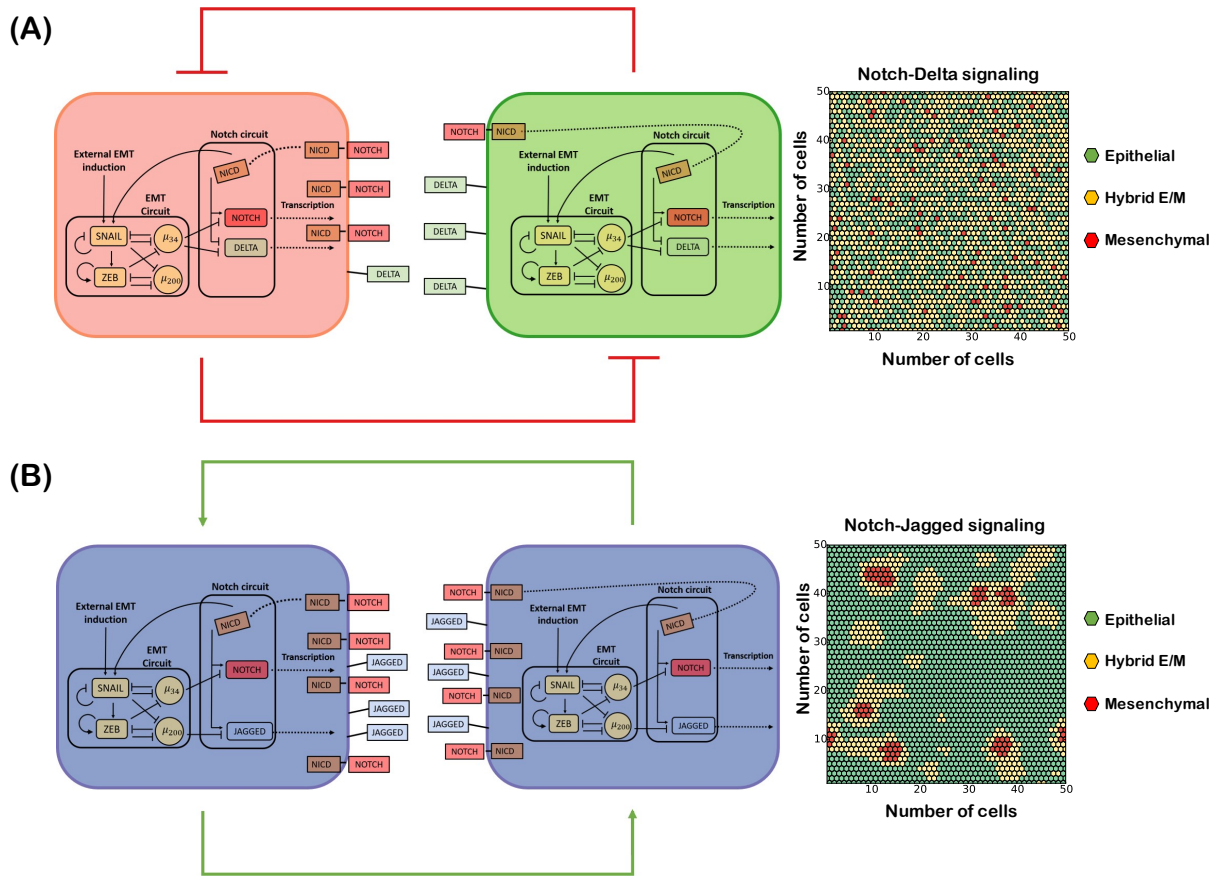

**Figure 4. Notch-Delta and Notch-Jagged signaling give rise to opposite cell patterning of EMT.** (A) Coupling of Notch-Delta signaling with the core EMT regulatory circuit. NICD suppresses the endogenous expression of Delta ligands upon activation of Notch receptors by exogenous Delta ligands (left, red color represents the 'receiver phenotype' (high notch), green color represents 'sender phenotype' (high delta)). Cell patterning of EMT in presence of a strong Notch-Delta signaling (right). (B) Coupling of Notch-Jagged signaling with the core EMT regulatory circuit. NICD promotes the endogenous expression of Jagged ligands upon activation of Notch receptors by exogenous Jagged ligands (left, blue color represents a 'sender/receiver phenotype' (high Jagged)). Cell patterning of EMT in presence of a strong Notch-Jagged signaling (right). Hexagons with different colors depict epithelial (green), hybrid E/M (yellow) and mesenchymal (red) cells. Figures in the right panel are adapted from [53].

The cell patterning mediated by Notch ligands can modulate epithelial-mesenchymal plasticity in a tumor tissue due to the aforementioned coupling between Notch and the EMT regulatory circuits. Intracellular Notch signaling activated by either Delta or Jagged can activate EMT. However, mathematical modeling of the coupled regulatory networks regulating EMT and Notch signaling suggests that Notch-Delta and Notch-Jagged signaling have dramatically different outcomes at a multi-cellular level. While Notch-Delta signaling promotes a spatial arrangement where cells in a partial or complete EMT are surrounded by epithelial cells, Notch-Jagged signaling can give rise to clusters of hybrid epithelial/mesenchymal (E/M) cells [74] (**Figure 4**). Indeed, Jagged1 is one of the top differentially expressed genes in collectively migrating cells of breast cancer [12,91]. These observations suggest that Jagged1 can act as an intercellular PSF that stabilizes a hybrid E/M phenotype in a non-cell autonomous manner. Therefore, in addition to 'conventional' PSF proteins that promote a partial EMT through direct crosstalk with the EMT regulatory circuitry, such as OVOL, GRHL2, and NRF2 [51,52,62], other PSFs can facilitate a partial EMT and the formation of CTC clusters by activating Notch-Jagged signaling and/or inhibiting Notch-Delta signaling. For example, NUMB/NUMBL that forms a negative feedback loop with Notch [92–94] can prevent a complete EMT, and consistent with that identification, knockdown of NUMB/NUMBL result in a mesenchymal phenotype and enables individual migration of the hybrid E/M NSCLC H1975 cells that typically migrate collectively [53].

### *Interaction among epithelial, hybrid E/M, and mesenchymal cells*
Similar to bidirectional interactions among epithelium and mesenchyme during organ development [95], there may be crosstalk and cooperation among cells exhibiting varying extents of EMT, which can accelerate tumor aggressiveness [96–98]. Hybrid E/M cells can facilitate such crosstalk, due to their plasticity that can generate and maintain epithelial-mesenchymal heterogeneity at a population level; such plasticity is limited for cells on either end of the EMT spectrum. Moreover, hybrid E/M cells can maintain cell-cell adhesion via E-cadherin, and thus potentially enabling formation of heterotypic clusters of CTCs with cells of varying EMT statuses [99]. Last but not least, a recent study highlighted that the hybrid E/M HMLER cells, but not a mixture of nonplastic 'extremely epithelial' (xE) cells and nonplastic 'extremely mesenchymal' (xM) cells, account for high tumorigenicity *in vivo* [100]. Notably, the nonplastic xE cells are created by ZEB1 knockdown and the nonplastic xM cells are created by constitutive ZEB1 expression, which supports the role of ZEB1 in mediating EMT [42,44].

Whether a mixture of E and M cells is sufficient to initiate the metastatic cascade and successfully form metastases remains to be resolved. Few molecular mechanisms enabling crosstalk between E and M cells have been elucidated recently. Tsuji *et al.* demonstrated that when EMT and non-EMT cells are inoculated subcutaneously in mice, they both establish primary tumors, but neither of them form lung metastases [96]; the ability to invade local tissues and enter circulation is demonstrated only by EMT cells. Further, when both cell types are injected intravenously, non-EMT cells form overt metastases, but the metastatic ability of EMT cells is compromised. Finally, when a mixture of EMT and non-EMT cells is subcutaneously injected, both cell types can intravasate but only non-EMT cells form lung metastases. Put together, these observations indicate a possible cooperation between EMT and non-EMT cells – while EMT cells can cleave the matrix to make way for both cell populations to intravasate, the non-EMT cells can colonize distant organs. This study did not identify any juxtacrine and/or paracrine signaling underlying this cooperation, but recent *in vitro* co-culture experiments of EMT and non-EMT cells have identified a few players that can mediate this crosstalk. HMLER cells overexpressing TWIST or SNAIL have been shown to impart migratory and invasive traits *in vitro* to control HMLER cells via paracrine Hedgehog signaling, but without explicitly inducing any morphological or molecular changes associated typically with EMT [97]. The authors also demonstrated that the EMT cells are able to increase the metastatic propensity of non-EMT cells *in vivo*, thus lending further credence to the notion that EMT cells can stimulate the migration of non-EMT cells. This idea is further strengthened by another *in vitro* analysis of sublines derived from prostate cancer PC-3 cells, where the relatively more epithelial PC-3/Mc cells, when co-cultured with post-EMT PC-3/S cells, have increased invasive potential which persisted for around 7 days after the co-culture but eventually decline after being segregated from PC-3/S cells [101]. This increased invasive response is also observed upon co-culture of PC-3/Mc cells with NIH3T3 fibroblasts, suggesting that the invasiveness of non-EMT cells can be increased by both tumor and non-tumor mesenchymal cells [101]. *In vivo* experiments for individual or co-injection of PC-3/S and/or PC-3/Mc cells corroborate previous observations that the post-EMT cells had little, if any, contribution to distant organ colonization [101]. Put together, these findings are reminiscent of *in vivo* studies showing that a persistent EMT activation can reduce metastases formation [6,102], and clinical evidence that carcinoma metastases are largely epithelial [103].

Reversible transitions among epithelial and mesenchymal phenotypes of disseminating cells has been dogmatically considered as the driving engine of metastasis for a long time [1,104,105] (also referred to as the 'sequential metastasis' model) [98], but with the proposed key role of EMT and MET being relooked at more carefully [20], the possibility of a cooperative journey taken together by epithelial and mesenchymal cells where they necessarily do not change their phenotypes cannot be ruled out (also referred to as the 'cooperative metastasis model') [98]. Multiple possibilities may underlie this cooperation: a) epithelial cells facilitate MET of mesenchymal cells, b) mesenchymal cells facilitate the survival, persistence and re-adhesion of epithelial cells during colonization, c) epithelial and mesenchymal cells exchange survival signals, d) a combination of the above. Nevertheless, collective transport of epithelial and mesenchymal cells is likely to be more effective for colonization as compared to that of one population alone. However, the role of hybrid E/M cells in this cooperative metastasis model remains to be explored.

What mechanisms may allow such collective migration of epithelial and mesenchymal cells? In developmental contexts, epithelial or mesenchymal cells have been seen to migrate collectively through respective cell-cell contacts mediated by E-cadherin or N-cadherin [106,107]. Recently, a N-cadherin/E-cadherin mechanically active heterophilic adhesion among the cancer-associated fibroblasts and cancer cells was reported to guide collective migration of tumor cells [108]. Given that the heterophilic E-cadherin/N-cadherin interaction has been proposed to be of similar affinity as that of homophilic E-cadherin interactions [109], collective cell migration can be expected to be observed among cells with varying levels of E-cadherin and N-cadherin. Another study reports a short-ranged interaction via EGF/CSF-1 paracrine axis to mediate macrophage-driven tumor cell migration [110,111]. CSF-1/CSF-1R axis has been recently proposed to associate with a hybrid E/M phenotype in inflammatory breast cancer [108] - a highly aggressive breast cancer subtype that metastasizes via clusters or emboli of circulating tumor cells [91]. These mechanisms may mediate, at least in part, a collective cooperative migration of cells in varying hues of EMT. Increased plasticity of hybrid E/M cells may be necessary and sufficient to maintain and propagate the non-genetic heterogeneity in terms of EMT status in a given isogenic cancer cell population.

**Mechanical control of EMT by ECM**

In addition to chemical communication, the stiffness of ECM also plays a key role in regulating EMT. For example, cancer cells when cultured in stiffer substrates exhibit increased migratory and invasive ability and become more mesenchymal-like [112]. Alternation in ECM stiffness can trigger multiple signaling pathways to regulate EMT, such as TWIST1-G3BP2 [23], HA-CD44 [113], MRTF-A [114], PI3K/Akt [115] and YAP/TAZ [116]. Yet, the reversibility of ECM stiffness-induced EMT can be cell line-dependent. For example, the mammary epithelial cells that have undergone EMT in a stiff substrate partially revert to epithelial phenotype [117], while the colon carcinoma HCT-8 cells can retain their mesenchymal-like phenotype, all after being re-cultured in the compliant substrate [118]. It is worth noting that cells undergoing EMT can in turn regulate ECM. For example, the LOX-family enzymes are upregulated in fibrosis and upregulation of LOX-family enzymes can directly increase connectivity of collagen fibers, stabilize and stiffen the collagen networks [119,120]. Given the importance of mechanical regulation of EMT, mathematical models that integrate mechanical with chemical signaling networks need to be developed to better understand EMT-ECM dynamics [121].

**Quantification of the EMT spectrum**

Our discussion so far has hopefully made it clear that cells undergoing EMT can acquire a spectrum of hybrid E/M states both *in vitro* and *in vivo*. However, the lack of a rigorous quantification of the EMT spectrum, namely, the exact proportions of epithelial, mesenchymal and hybrid E/M subpopulations of cell lines and clinical samples, can lead to potentially contradictory conclusions regarding the necessity and functional roles of EMT and MET in metastasis [15–17,102,122,123].

Cells undergoing EMT typically alter both their omics profiles and morphologies. Therefore, in principle, the EMT spectrum can be quantified via evaluating the change of cell morphology and/or their omics profiles. To classify epithelial and mesenchymal phenotype at a single-cell level, Leggett *et al.* developed a probabilistic classification scheme using Gaussian mixture model (GMM) focusing on 4 morphological features of single cells – maximum radius of the nucleus, vimentin area, cytoplasm form factor and maximum feret diameter of cytoplasm [124]. The GMM is trained using the morphological features of DMSO-treated (epithelial) and 4-hydroxytamoxifen (OHT)-treated (mesenchymal) human mammary MCF-10A cells which are transfected with an inducible Snail construct, referred to as MCF-10A Snail cells. The probabilistic GMM model has revealed various EMT kinetics of MCF-10A Snail cells when induced by TGF-β1, plating density and the microtubule inhibitor Taxol respectively. This GMM model also provides insight into the EMT status of individual cells which may be overlooked by population-average analysis. However, this method only focuses on a binary classification of EMT – epithelial or mesenchymal and one missing piece of this model is the classification of hybrid E/M phenotypes.

To quantitatively measure the EMT status of cell lines with specific attention to a hybrid E/M phenotype, George *et al.* developed an EMT scoring metric to calculate the probability of a given sample to be hybrid E/M phenotype and assign a score between 0 and 2 with 0 being fully epithelial, 2 being fully mesenchymal and 1 representing hybrid E/M [125]. Using the gene expression data of NCI-60 human tumor cell lines as the training set, the ratio of VIM to CDH1 together with CLDN7 expression are identified as the best-fit pair of predictors to classify EMT phenotypes. This EMT scoring metric has been used to characterize multiple hybrid E/M cancer cell lines including A549 and DU145. Furthermore, this EMT scoring metric has been extended to distinguish hybrid E/M cells from mixtures of epithelial and mesenchymal cells [60]. Another EMT scoring metric developed by Tan *et al.* assigns a score between -1 and 1 to a given sample with -1 being fully epithelial and 1 being fully mesenchymal [126]. Both George *et al.* and Tan *et al.* demonstrated that patient samples that are more mesenchymal-like do not necessarily correlate with worse overall and disease-free survival results and do not always show resistance to chemotherapy, indicating a subtype-dependent role of EMT in cancer progression and therapy resistance. In summary, these methods to quantify EMT status help address the multifaced roles of EMT in tumor progression and patient prognosis.

**EMT and CTC clusters**

Cancer cells that acquire a hybrid E/M phenotype maintain both epithelial (e.g. cell-cell adhesion) and mesenchymal (e.g. migration) features thus can migrate collectively as a cluster. Such clusters of migrating tumor cells have been shown to be one of the primary instigators of metastases [12,127]. The experimental studies supporting this notion are discussed below.

Using multicolor lineage tracing, Cheung *et al.* showed by two sets of experiments that the CTC clusters are mainly formed by multi-cellular clusters from the primary tumors. In the first set of experiments, two differently colored populations - mTomato+ and CFP+ MMTV-PyMT tumor organoids were respectively transplanted into the right and left flanks of a nonfluorescent mouse. After 6 to 8 weeks, only single-colored metastases were observed in lung. In the second set of experiments, mTomato+ single cells isolated by FACS were transplanted into a nonfluorescent mouse via tail-vein injection and 2 days later FACS isolated CFP+ single cells were injected into the same mouse. After 2 days, exclusively single-colored metastases were observed in lung. These two sets of experiments suggest that the polyclonal metastases in lung is more effectively generated by multicellular seeds and not by serial aggregation of single tumor cells [12]. The CTC clusters exhibit enriched expression of an epithelial cytoskeletal protein K14 that is required for the collective invasive behavior and distant metastasis. Another study supporting the idea that CTC clusters can arise as oligoclonal groups of cells detached from the primary tumors is Aceto *et al.* who identified plakoglobin as a key mediator for CTC cluster formation [128].

An alternative mechanism – that CTC clusters can be formed via aggregation of single CTCs in circulation - has been recently demonstrated by Liu *et al.* using intravital multiphoton microscopic imaging in both patient-derived xenograft (PDX) models of triple negative breast cancer (TNBC) and PyMT transgenic mouse models [127]. Liu *et al.* first showed that in chemoattractant-containing matrigel about 20% of invasion events of the tumor cells collected from the PDX models occur as multicellular aggregates, and in suspension culture individual CTCs derived from a breast cancer patient aggregate into clusters within 1 to 2 hours. These results indicate that single CTCs can aggregate *in vitro*. The authors further co-infused eGFP+ and tdTomato+ single MDA-MB-231 cells via the tail vein and observed that about 92% of lung metastasis are dual-color aggregates within 2 hours. The percentage of the dual-colored aggregates is affected by the timing of the eGFP$^+$ and tdTomato$^+$ entering blood vessels. Consequently, sequential infusion of the tdTomato$^+$ cells 5 minutes, 10 minutes and 2 hours after the infusion of the eGFP$^+$ cells result in gradually decreased percentages (27%, 16% and 10%) of the dual-colored aggregates. The authors further showed that the CTC clusters exhibit notably enriched expression of CD44, and that the intercellular CD44 homophilic interaction is responsible for the aggregation of single CTCs.

One major difference in the studies by Cheung *et al.* and Liu *et al.* is the timing of the second-colored single cells entering the blood vessels. Cheung *et al.* injected the second-colored cells into the tail vein 2 days after the injection of the first-colored cells while Liu *et al.* waited for at most 2 hours to inject the second-colored cells. As discussed by Liu *et al.*, the timing of the second-colored cells entering the blood vessel could have significant effect on the percentages of dual-colored lung metastases. From our perspective, the key issue in these two studies seems to be the lifetime and density of the injected single cancer cells. Since it is not clear how often the injected cells are expected to reach the bloodstream, it is hard to evaluate which protocol is a better match to reality. The experiments by Cheung *et al.* where two differently colored populations of tumor organoids are injected respectively into the right and left flanks of nonfluorescent mice seems to be a better approach since the interaction of cells in the bloodstream is determined naturally rather than by the experimenters. Nonetheless, both studies showed that the CTC clusters significantly promote colony formation and lung polyclonal metastases *in vivo*. Liu *et al.* showed through CellSearch platform-based blood analysis that the breast cancer patients with CTC clusters exhibit significantly worse overall survival results relative to those with only single CTCs. And, CTC clusters in different contexts can be different. For example, Khalil *et al.* showed that though invasive ductal carcinoma (IDC) and invasive lobular carcinoma (ILC) exhibit collective invasion patterns, the collective invasion in IDC lesions maintains intercellular E-cadherin while collective invasion in ILC lesions loses intercellular E-cadherin but retains CD44 for intercellular junctions [129]. All together, these studies suggest that CTC clusters can in principle be generated both by cohesive shedding from the primary tumors and/or serial aggregation of single cells in the circulation.

To model the migration of CTC clusters, Bocci *et al.* proposed a simple biophysical model where cancer cells can undergo a partial EMT that allows both single-cell and clustered-cell migration, or a complete EMT that only allows single-cell migration [130]. According to this reduced physical model, a tumor can undergo a transition from primarily single cell-based invasion to collective invasion. Strikingly, this theoretical framework reproduces multiple CTC cluster size distributions measured in patients and mouse models across cancer types, hence suggesting the existence of a unifying set of principles governing cancer cell migration [130].

Adding to the complexity of the problem, CTC clusters can associate with non-cancer cells including platelets and/or immune cells such as tumor-associated macrophages and neutrophils [131–133]. In particular, it is well recognized that interactions between macrophages and tumor cells are very important for tumor cell intravasation and extravasation, though the detailed mechanisms have not been elucidated [134,135]. In addition, the CTC-neutrophil clusters lead to more efficient metastasis relative to CTCs alone [131]. Future studies need to be performed to elucidate the cross-talk between CTC clusters and different types of immune cells.

**EMT and stemness in tumor microenvironment**

We have already come across the fact that besides the migratory and invasive traits conferred by EMT, some cancer cells undergoing EMT can also acquire an enhanced ability to drive tumor initiation and an enhanced therapy resistance. These properties are associated with the notion of 'stemness', and such cells are sometimes referred to as cancer stem cells (CSCs). The connection between EMT and stemness was first proposed by Brabletz *et al.* [24] more than a decade ago based on the premise that EMT and stemness cannot explain the different steps of the metastatic cascade when considered as separate and independent processes [24]. Indeed, increasing experimental evidence suggests a strong association of EMT with stemness. For instance, human mammary epithelial cells (HMLEs) undergoing EMT can express stem cell markers and exhibit increased mammosphere formation [122]. Moreover, reversing EMT via knockdown of SNAIL represses stemness and tumor growth in ovarian cancer [136]. This and other evidence suggests that the activation of the EMT program leads to the acquisition of stemness [137]. This correlation, however, is not absolute, and several other experimental papers have claimed that stemness can occasionally correlate with an epithelial phenotype or suppression of EMT [101,102,138].

Revisiting the EMT-CSC connection through the lenses of epithelial-mesenchymal plasticity and hybrid E/M phenotypes, however, offers a more consistent picture. Mathematical modeling approaches have been applied to decode the connection of EMT with stemness via simulating the coupled decision-making regulatory networks of EMT and stemness. Specifically, EMT activation can downregulate let-7, a miRNA typically associated with repressing stemness. Let-7 can bind to ZEB and promote its degradation. These two processes can settle at an intermediate level of let-7 and ZEB leading to hybrid E/M CSCs [139]. This association, however, can vary depending on the coupling between the modules governing EMT and stemness, allowing for CSCs in an epithelial, mesenchymal or hybrid E/M phenotype [139]. In addition, local perturbations in the tumor microenvironment such as TGF-β or Notch signaling also can modulate the association of CSCs with different EMT phenotypes [140,141]. All told, emerging evidence from theoretical and experimental studies tends to associate the hybrid epithelial/mesenchymal phenotypes with stemness [139,140,142–145].

Along these lines, the association of hybrid E/M phenotypes with stemness can be promoted by the PSFs – OVOL and Jagged1. Indeed, Jagged1 is typically overexpressed in CSC populations as compared with non-CSCs in multiple types of cancer including glioblastoma, pancreatic cancer, colon cancer, and breast cancer [22,146–148]. Through modeling the coupling of the regulatory networks governing EMT, stemness and Notch signaling, Bocci *et al.* proposed that Jagged1 facilitates a 'window of opportunity' that confers maximal invasion potential in terms of EMT and stemness [140,145]. A follow-up work showed that knockdown of Jagged1 impairs breast organoid formation *in vitro*, therefore highlighting a causal relationship between Jagged1 and stemness [141].

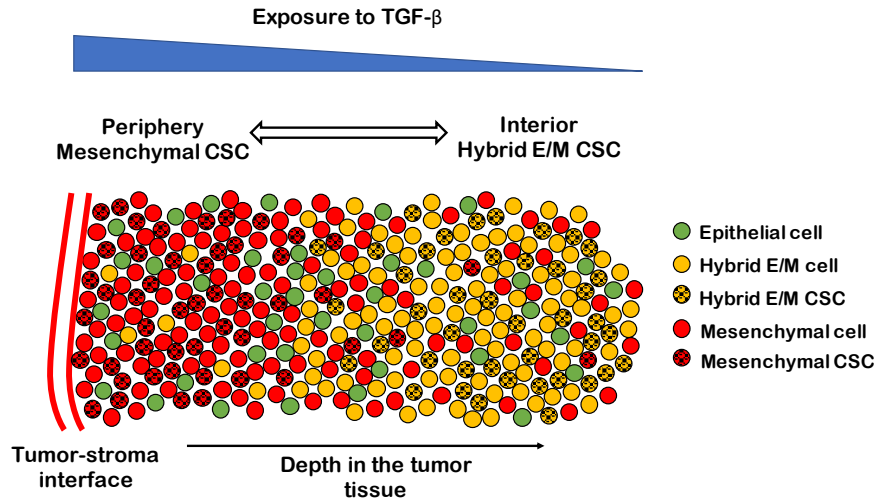

**Figure 5. The spatial patterning of CSCs with different EMT phenotypes.** Tumor-stroma interactions can give rise to a gradient of TGF- β (top, blue scale). In the periphery of the tumor, most cells are mesenchymal (red spheres), while the interior is mostly composed by hybrid E/M and epithelial cells (yellow and green spheres, respectively). CSCs are mostly mesenchymal in the periphery (black-dotted red spheres) and mostly hybrid E/M in the interior (black-dotted yellow spheres).

Subpopulations of CSCs with different EMT phenotypes can be representative of the spatial organization of a tumor tissue. In breast cancer, $CD24^-/CD44^+$ mesenchymal CSCs are typically located toward the invasive edge of the tumor by the tumor-stroma interface, while $ALDH1^+$ CSCs are found in the more interior region [149]. Interestingly, $ALDH1^+$ CSCs were originally considered as epithelial-like, but their RNA sequencing (RNA-Seq) data has later shown that these cells lean more toward a hybrid E/M phenotype and share several genes with TNBC signature [150]. Recently, Bocci *et al.* have argued that this spatial segregation can be qualitatively explained by the interplay of cancer cells with the tumor microenvironment [141]. Cytokines such as the EMT-inducer TGF-β are typically secreted at the tumor-stroma interface and give rise to a gradient of EMT-inducing signal throughout the tumor tissue. Therefore, CSCs at the invasive edge are highly exposed to TGF-β and tend to undergo a complete EMT leading to a fully mesenchymal phenotype, while CSCs in the interior maintain a hybrid E/M phenotype [141] (**Figure 5**). Consistently, in non-small cell lung CSCs, a subpopulation of mesenchymal CSCs exhibits high expression of TGF-β targets such as SNAI1 and ZEB1, as compared to a hybrid E/M CSC population [140].

In summary, EMT and CSC formation represent two essential axes of tumor progression, whose connection is modulated by factors including tumor microenvironment and cell-cell signaling [141,149,151,152]. The stem cell paradigm inherited from developmental biology implies a hierarchical lineage of cells that gradually but irreversibly differentiate [153,154]. In the context of cancer, however, stemness has proven to be a dynamical property that cells can gain and lose [155–157]. The complex interplay between EMT, stemness and tumor microenvironment gives rise to tumor heterogeneity that still represents the major challenge hindering metastasis and therapy resistance [151].

**EMT and immune suppression**

In addition to cancer cells, a solid tumor harbors other types of cells which form the tumor microenvironment and strongly affects cancer outcome [26]. Specifically, many groups have investigated the roles of immune cells in cancer progression. Certain types of immune cells, such as macrophages and T cells, can comprise up to 50% of the cells in a solid tumor [27]. These immune cells usually polarize into phenotypes, such as M2-like macrophages and regulatory T helper cells (Tregs), that promote tumor progression via suppressing the activity or viability of anti-cancerous immune cells, e.g., M1-like macrophages and cancer-killing T effector cells [26,27,158]. Our goal here is to focus on the role of EMT in this tumor-immune interplay.

A series of mathematical models have been proposed to understand the roles of tumor-immune interactions in polarizing the cytokine-immune cell network into states dominated by either immune-promoting or immune-suppressing populations [159–164]. Such modeling frameworks can help to design effective perturbations to revert the immune microenvironment from an immune-suppressing to an immune-promoting one. Many of these models consider the fact that macrophages and cancer cells can directly interact with each other and regulate the behaviors of one another, as shown by many *in vitro* experiments [165–170]. The interactions between macrophages and cancer cells are formidably complex, and the emergent dynamics can be non-intuitive. A series of mathematical models capturing these interactions suggests that cancer cells in the epithelial-like state (less aggressive) and M1-like macrophages might form a stable ecosystem, whereas cancer cells in the mesenchymal-like state (more aggressive) and M2-like macrophages form another stable pair [171].

The question of establishing an immune-dominated versus immune-suppressed microenvironment should have an effect on the activity of cd8+ effector T-cells. There have been studies suggesting that T cells tend to be excluded from tumors enriched by mesenchymal-like cancer cells [172,173]; we will discuss this further below. Combining this data with the modeling results, one could argue that interactions between macrophages and cancer cells drive the macrophages to be M2-like and the cancer cells to be mesenchymal-like; and due to the effects of M2-like macrophages, T cells may be excluded from tumor areas enriched with mesenchymal-like cancer cells. Since both the infiltration of cancer-killing immune cells [33–39] and the epithelial-mesenchymal plasticity of cancer cells are important for the prognosis of cancer patients, it is clearly valuable to evaluate the association and ultimately the casual relationship between the two.

In the following, we will first describe in greater detail the corresponding *in vitro* and *in vivo* experiments as well as analyses of gene expression data from The Cancer Genome Atlas (TCGA) (**Figure 6**). Finally, we discuss the potential causal relationship between the EMT status of cancer cells and the infiltration of cancer-killing immune cells.

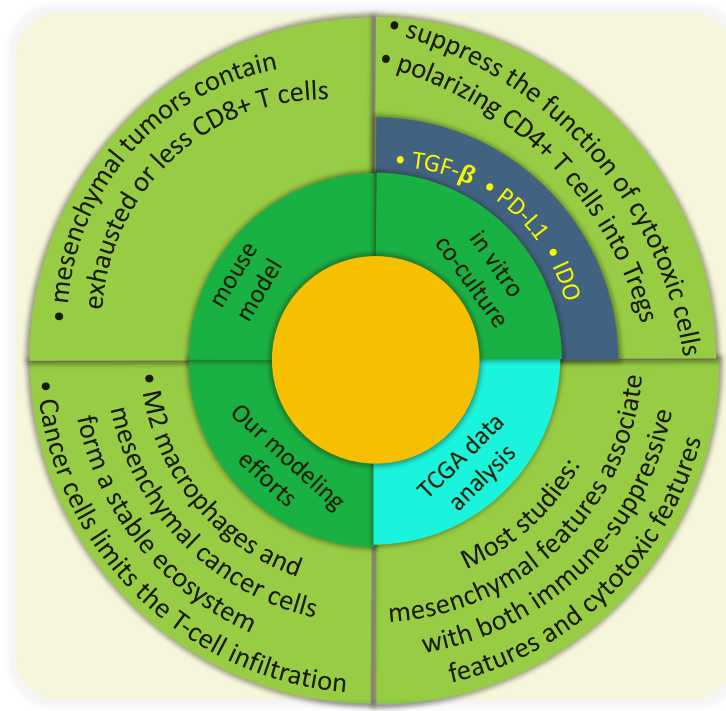

**Figure 6. The relationship between EMT and immune response as shown by various approaches.**

*In vitro characterization of the immune-suppressing role of mesenchymal-like cancer cells*

For the interaction between mesenchymal-like cancer cells and immune cells, TGF-β signaling has been studied extensively. TGF-β is a well-known EMT inducer [21] and can be secreted by tumor-associated fibroblasts [174,175] and tumor-associated macrophages [176]. TGF-β signaling can impair maturation, differentiation, and/or activation of both innate and adaptive immune cells [31,177]. Specifically, TGF-β can inhibit the functions of cytotoxic T-cell functions [178,179]. In the co-culture experiments, Joffroy *et al.* showed that TGF-β secreted by cancer cells induces the differentiation of $CD4^+$ T cells into Treg ($Foxp3^+$) cells, which are immune-modulating cells [178]. Similar cancer-cell dependent expansion of $CD4^+Foxp3^+$ T cells is also shown by Kudo-Saito *et al.* when co-culturing SNAIL-enhanced mouse melanoma B16F10 cells with $CD4^+$ T cells [172]. Furthermore, TGF-β can downregulate the MHC class I proteins as shown in prostate cancer [180] as well as NSCLC cell lines [181].

Aside from TGF-β, the PD-L1/PD-1 axis has also attracted attentions due to its implications in the immune-checkpoint blockade therapy. Presumably, PD-L1 expressed by cancer cells can bind to PD-1 expressed on the surface of cancer-killing immune cells, which will modulate their cytotoxic functions. It is shown that when driving EMT via downregulating miR-200 and overexpressing ZEB1, PD-L1 expression by cancer cells is upregulated [182]. Therefore, mesenchymal-like cancer cells could be more resistant to cancer-killing immune cells by upregulating PD-L1. Interestingly, expression levels of both TGF-β1 and PD-L1 by cancer cells can be induced under hypoxic conditions [183–185]. The hypoxia-induced EMT can promote immunosuppression via induced expression of indoleamine 2, 3-dioxygenase (IDO, another T-cell suppressing factor) in monocyte-derived macrophages [186].

EMT can also be induced when cancer cells are challenged by immune cells or inflammatory cytokines [187]. In addition, cancer cells can be equipped with immunomodulatory effects which interfere with proliferation, differentiation and apoptosis of NK, T-cell, and B-cell populations [187]. As was the case for hypoxia, the IDO pathway is important for the immunomodulatory effects on T cells after inflammation-induced EMT [187]. These experimental results are potentially helpful for establishing a potential causal relationship between EMT of cancer cells and the infiltration of cytotoxic immune cells.

*In vivo characterization of the immune-suppressing role of mesenchymal-like cancer cells*

Generally, *in vivo* experiments using mouse models show that tumors formed by mesenchymal-like cancer cells are less infiltrated with cancer-killing immune cells and/or more infiltrated by immune-suppressing immune cells. For example, Kudo-Saito *et al.* showed that in mice, compared with tumors formed by mock-transfected B16F10 cells, tumors derived from Snail-transduced B16F10 cells exhibit less tumor-infiltrating $CD8^+$ T cells, while more Tregs, and form more lung metastases [172]. In the same mouse model, the chemokine CCL2 can recruit immune-modulating populations such as MDSC and macrophages [188], which are responsible for creating the immune-suppressing microenvironment. In addition, in a mouse model of breast cancer [173], tumors formed by mesenchymal carcinoma cell lines contained more Tregs, M2 macrophages and exhausted $CD8^+$ T cells, whereas tumors formed by more epithelial carcinoma cell lines contained $CD8^+$ T cells and M1 macrophages. Furthermore, using transgenic mouse models, Spranger *et al.* showed that an active β-catenin signaling in cancer cells contributes to a lack of T cell infiltration in tumor sites and resistance to anti-PD-L1 and/or anti-CTLA4 mAb therapy [189].

In summary, from the perspective of direct experiment *in vitro* or in preclinical models it is becoming clear that mesenchymal-like cancer cells can directly suppress the function of cancer-killing immune cells as well as promote the immune-suppressing microenvironment by recruiting or polarizing immune-suppressing immune cells. Following this logic, one would expect a lower presence of functional cancer-killing immune cells in the tumor area enriched with mesenchymal-like cancer cells.

*Gene expression data analysis*

What about the results for patients? Unfortunately, the overall picture here as it emerges from TCGA data analyses tends to suggest that while mesenchymal tumors are generally enriched with immune-suppressing cells, they are often enriched with cancer-killing immune cells as well, as compared with epithelial tumors. For example, Mak *et al.* showed that the pan-cancer tumors samples with high EMT scores correlate with high expression of several immune checkpoints including PD-1, PD-L1, CTLA4, OX40L, and PD-L2 [190]. Lou *et al.* observed similar trend in lung cancer patients with early or advanced NSCLC adenocarcinomas [191]. Aside from the immune checkpoint markers, Lou *et al.* also found that the lung tumors that displayed an EMT phenotype also have a higher infiltration of Tregs. Interestingly, in their work, some immune costimulatory molecules such as CD80 and CD86 as well as IFN-γ signals are more highly expressed in "mesenchymal" lung adenocarcinoma. Reports also suggest that, in the claudin-low subtype of breast and bladder cancer, which are mesenchymal-like, tumors are generally well-inflamed by immune-promoting immune cells but these cells are under active immunosuppression [192].

It should be noted that data in TCGA is rarely from tumor cells exclusively, thus the mesenchymal features seen there may be a consequence of higher infiltration of stromal cells. Conversely, analysis of epithelial markers [193] may not be biased by stromal cells. When investigating the epithelial marker ESRP1 for melanoma samples in TCGA dataset, Yao *et al.* found that a high infiltrating lymphocyte activity is enriched in ESRP1-low melanoma samples which have enhanced mesenchymal features [193]. The lymphocyte activity was evaluated by the gene expression of two cytotoxic agents - perforin (PRF1) and granzyme A (GZMA). Considering the use of an epithelial marker instead of mesenchymal markers, this work may be a strong piece of evidence, since the contamination by mesenchymal markers from non-cancer cells is supposed to be low. However, the immune infiltration still needs to be defined rigorously. It is possible that cytotoxic immune cells are constrained to lie in the tumor stroma instead of infiltrating into the tumor islets, though many of them infiltrate into the core of the tumor. For these tumors, the bulk tumor gene expression data will give a high infiltration score of cytotoxic immune cells. If these tumors happen to be ESRP1-low tumors, we are likely to conclude that a higher infiltration of cytotoxic immune cells is associated with more mesenchymal-like cancer cells.

Although the above-mentioned evidence tends to suggest that a higher infiltration of cytotoxic immune cells associates with the mesenchymal features of tumor samples, the jury is still out, and an opposite trend has been reported elsewhere. For example, Chae *et al.* reported that EMT is associated with significantly lower infiltration of CD4/CD8 T cells in squamous cell carcinoma [194]. In addition, high numbers of CD8$^+$ TILs have been shown to correlate with low SNAIL expression in extrahepatic cholangiocarcinoma [195]. Furthermore, in bladder cancers, patients with tumors characterized by an epithelial (luminal) phenotype have a better response rate when treated with anti-PD-L1 therapy compared to those harboring basal subtype [196], which are mesenchymal-like [197]. Since it has been suggested that patients who respond to anti-PD-L1 therapy usually have pre-existed CD8$^+$ T cells [198] which can be unleashed by the therapy, it is likely that in this particular study, epithelial-like tumors are better infiltrated by CD8$^+$ T cells than mesenchymal-like tumors. In addition, in muscle-invasive urothelial bladder cancer (MIUBC), activation of β-catenin pathway is found in the most common non-T cell-inflamed MIUBCs [199].

In summary, most of the gene expression analyses point towards a positive correlation between mesenchymal-like cancer cells and the higher infiltration of both immune-promoting and immune-suppressing immune cells, which seems the opposite of the trend revealed by *in vitro* co-culture experiments and *in vivo* mouse models. The inconsistency can be simply due to the fact that different analyses use varying standards of assessing the abundance of cancer-killing immune cells, for example, gene expression signatures of cancer-killing immune cells [189] or the percentage of cancer-killing immune cells among all (immune) cells [173]. In addition, the selected regions of interest (ROI) can also be inconsistent - core vs. margin – with varying immune-associated traits. Furthermore, the number of cancer cells in different ROI need to be evaluated for a fair comparison,, i.e., a particular ROI can be less inflamed. This confusion could be at least partially resolved if the infiltration could be quantified on the tumor islets level and the EMT status determined for single cancer cells instead of for the bulk population. For example, Seo *et al.* stained both EMT and CD8 markers for one breast cancer patient and found a positive correlation between the two [200]. However, the markers were not on the same section and there is still an issue of defining the immune infiltration. Therefore, it would be ideal to study the gene expression/protein abundance of the tumor islets or the sorted tumor cells in addition to the analysis of the corresponding images of the same tumor, so that we can have a better idea about whether there is any association between the EMT status of cancer cells and the infiltration of cancer-killing immune cells into the tumor islets.

***Causal relationship between EMT and infiltration of cancer-killing immune cells***

Even if we overcome the above-mentioned issues regarding the correlation between EMT and the infiltration of cancer-killing immune cells, there is still a hard problem to understand the causal relationship between the two. There is here an issue of the temporal coevolution of the tumor microenvironment which could play a role. Specifically, it is possible that apparently contradictory observations are in fact different snapshots of the co-evolution process of the two sets of cells. We can propose two potential scenarios on the causal relationship between the EMT status of cancer cells and the polarity of the immune microenvironment as well as the infiltration of cancer-killing immune cells.

The first scenario: tumor regions enriched with epithelial cancer cells first attract cancer-killing immune cells and then cancer cells are attacked by these immune cells. If one investigates the infiltration of immune cells at this stage, these epithelial tumors should be enriched by cancer-killing immune cells. The immune attack may convert epithelial-like cancer cells into mesenchymal-like ones [187]. Then the mesenchymal-like cancer cells can suppress the function of cancer-killing immune cells and start to promote the accumulation of immune-suppressing immune cells. Following this logic, at this stage, the mesenchymal-enriched tumors regions should have the infiltration of both cancer-killing and immune-suppressing immune cells. The TCGA data analysis results seem to support this hypothesis [190–192]. However, many *in vivo* experiments do not seem to support the late stage assumption of this scenario [172,189] even though experiments using epithelial-like cancer cell lines tend to support the early state aspects of this scenario at the early state [173].

The second scenario: before the engagement of immune systems, the cancer cells interact mainly with the cancer-associated fibroblasts (CAFs) or the tumor-associated macrophages (TAMs). The paracrine signaling between the two (cancer cells and CAFs, or cancer cells and TAMs) biases the cancer cells towards mesenchymal-like phenotypes [174–176]. Consequently, those mesenchymal-like cancer cells tend to exclude cancer-killing immune cells from the microenvironment. This exclusion may be also due to the accumulation of TAMs [188] or the absence of specific types of dendritic cells [189]. In addition, the mesenchymal-like cancer cells may also have their own ways to stop or exclude cancer-killing T cells [178,179,182]. In this scenario, epithelial tumors that do not undergo EMT (due to interactions with CAFs or TAMs) may be infiltrated by cancer-killing immune cells whereas mesenchymal tumors tend to exclude cancer-killing immune cells. Apparently, this scenario seems to be supported by the *in vitro* co-culture experiments as well as some *in vivo* studies but is not supported by most TCGA data analyses results.

In order to test different scenarios, it will be essential to monitor the trajectory of EMT of cancer cells as well as the evolution of the infiltration of cancer-killing immune cells in a single mouse. It has been shown that even in a single mouse model, some tumors can be "hot", and others can be "cold" [201]. The particular mouse model used in this recent study can be an ideal system to study the co-evolution of the EMT status of cancer cells as well as the infiltration pattern of cancer-killing immune cells to better understand the causal relationship between the two.

**Hybrid E/M phenotypes and beyond**

Our claim here has been that hybrid E/M phenotypes enable cancer cells metastatic plasticity to effectively metastasize. Notably, such stable hybrid phenotypes that combines features of two speciously exclusive phenotypes is not limited to EMT. For example, during pathological angiogenesis in cancer, a stable hybrid tip/stalk phenotype that results in poorly perfused and chaotic angiogenesis can be acquired due to the overexpressed Jagged [202]. The elevated production of Jagged in cancer can also promote the existence of a stable hybrid sender/receiver state where cells maintain intermediate levels of Jagged and Delta (ligands) and Notch (receptor) thus allowing cells the plasticity to both send and receive signals [90]. Another example of a hybrid phenotype in cancer appears to arise in cancer metabolic preprogramming. Some cancer cells can acquire a stable hybrid metabolic phenotype where both glycolysis and oxidative phosphorylation (OXPHOS) can be utilized [203–205]. Such hybrid metabolic phenotype enables cancer cell metabolic plasticity to grow and prosper in various hostile microenvironments. The hybrid metabolic phenotype has been observed in the human TNBC SUM159-PT and MDA-MB-231 and mouse breast cancer 4T1 cells at least at a population level. The emergence of the hybrid metabolic phenotype indicates that targeting both glycolysis and OXPHOS may be necessary to eliminate cancer metabolic plasticity. Notably, such a hybrid metabolic phenotype is not limited to cancer cells. For example, the naive pluripotent stem cells (PSCs) exhibit a hybrid glycolysis/OXPHOS phenotype relative to primed PSCs which primarily use glycolysis and somatic cells which primarily use OXPHOS [206]. As a result, activation of both OXPHOS and glycolysis, synergistically by the TFs Zic3 and Esrrb for example, is essential to reprogram somatic cells to acquire the naive pluripotency. In addition, regulatory T cells that utilize both glycolysis and fatty acid oxidation can be more effective in expansion than those that primarily use glycolysis [207]. In our final example, the T-helper (Th) cells can acquire a stable hybrid Th1/Th2 phenotype where cells co-express two mutually inhibiting TFs T-bet (stimulating Th1) and GATA3 (stimulating Th2) [208]. The hybrid Th1/Th2 phenotype combines the properties of Th1 and Th2 and can regulate effector T cell to function without excessive inflammation. In summary, it is reasonable to speculate that the hybrid phenotype is a common characteristic emerging from cellular plasticity.

**Conclusions and Future Vision**

Epithelial-mesenchymal plasticity has attracted much attention due to its critical roles in facilitating metastases, stemness and immune repression. Emerging evidence suggests that cancer cells can acquire a spectrum of hybrid E/M phenotypes and cells can transition back and forth between different EMT phenotypes. Computational and experimental studies have been combined to deepen our understanding of EMT dynamics, the emergence of hybrid E/M phenotypes and the interplay of EMT with stemness and immune suppression. Notably, EMT is a multi-dimensional spatiotemporal program including alterations in mRNA and protein abundance, protein modification and relocation which can mediate the changes in cell junction, morphology and apico-basal polarity, resulting in adoption of cell migratory and invasive properties. Future computational studies need to integrate different facets of EMT to systemically elucidate EMT dynamics and quantify EMT status. The advance in experimental technologies such as simultaneous measurement of transcriptome and proteome at a single-cell level - CITE-seq [209] and REAP-seq [210], intravital correlative microscopy [211] will contribute to high-resolution quantification of EMT *in vivo*. The combination of theoretical and experimental efforts will continue uncovering important mechanisms underlying epithelial-mesenchymal plasticity and its association with other hallmarks of cancer.

Thus, the emergence of hybrid phenotypes in cancer and the ability of cancer cells to switch back and forth between various phenotypes indicate that therapeutic strategies targeting cancer cell plasticity needs to be designed more carefully. Eliminating the notoriously aggressive hybrid phenotypes in cancer may serve as the first step towards conquering cancer.

**Acknowledgements:**

This work is supported by the National Science Foundation (NSF) grant for the Center for Theoretical Biological Physics NSF PHY-1427654, by NSF grants PHY-1605817 and CHE-1614101 and by the John S. Dunn Foundation Collaborative Research Award. MKJ was supported by Ramanujan Fellowship awarded by SERB, DST, Government of India (SB/S2/RJN-049/2018). FB was supported by the Marjory Meyer Hasselmann Fellowship for academic excellence in chemistry.